\documentclass[aps,manuscript]{revtex4}

\usepackage{graphicx}

\begin{document}

\title{Fluorescence interferometry}

\author{G.~S.~Paraoanu}\email{paraoanu@cc.hut.fi}
\affiliation{Low Temperature Laboratory, Aalto University, P. O. Box 15100, FI-00076 AALTO, Finland.}

\begin{abstract}
We describe an interferometer based on fluorescent emission of radiation of two qubits in quasi-one-dimensional modes. Such a system can be readily realized with dipole emitters near conducting surface-plasmonic nanowires or with superconducting qubits coupled to coplanar waveguide transmission lines.
\end{abstract}

\pacs{07.60.Ly,03.70.+k,03.65.-w}
\maketitle

\section{Introduction}

Interferometers have played a fundamental role in the development of experimental and theoretical physics. The classic two-slit interference experiments, the Michelson-Morley interferometer, the Mach-Zender interferometer, the HOM (Hong-Ou-Mandel) two-photon interferometer, all these comprise an essential set of tools in the field of quantum optics \cite{optics}. In this paper, we show that it is possible to construct a new type of interferometer in which the beam of an incoming field is split by a qubit (we will generically use this term to cover any dipolar-coupled two level system) and recombined by another one. This instrument can serve for detecting the phase difference created by an object (or in general due to any optical path difference) placed in the path of the field propagating in one of the arms of the device. This proposal is motivated by the newest experimental advances in the field of plasmonics and superconducting circuits. Superconducting two-qubit systems driven by fields have attracted a lot of interest due  to their potential applications in quantum information processing (see {\it e.g.} \cite{me} and references therein), and strong coupling between superconducting qubits and coplanar waveguide transmission lines has been achieved \cite{circuitqed}. Recently, the resonance fluorescence spectrum of a flux qubit into such a line has been measured \cite{fluorescence}. Strong coupling between molecules and surface plasmons is already an experimental reality \cite{paivi}, and devices such as single-photon transistors are possible with the use of nanowires \cite{lukin}. In such systems, the main relaxation channel for the qubit  is provided by the
quasi one-dimensional modes (nanowires, coplanar waveguides), which can be engineered such that large Purcell factors are obtained \cite{circuitqed,lukin}.

\section{General quantum network}

\begin{figure}[t]
\begin{center}
  \includegraphics[width=13cm]{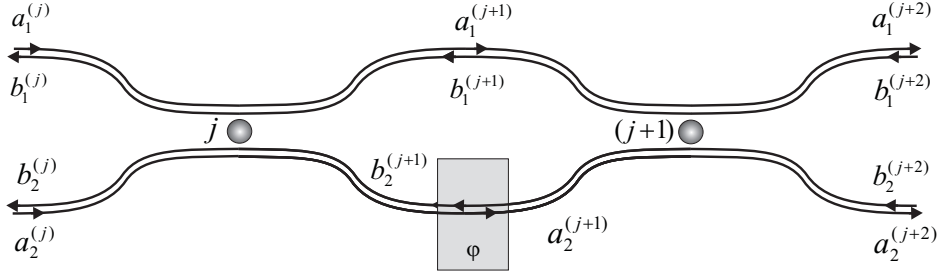}
\end{center}
\caption{Schematic of the quantum network. Only two modes, 1 and 2, are  illustrated for simplicity. The qubits, indexed by $j$, are placed at positions $x^{(j)}$, and the field propagating in mode 2 acquires an additional phase difference $\varphi$.}
         \label{schematic}
\end{figure}

The system we consider consists of qubits coupled to different modes of the electromagnetic field (Fig. \ref{schematic}), with the possibility of adding a phase shifts in one of the modes ({\it e.g.} by placing a material, another off-resonant qubit, or by simply having a longer path). Although our final results are for the case of two qubits and two modes (which constitute an interferometer), we show that one can give an analytically exact description of such a quantum network in the most  general case, namely $M$ quasi one-dimensional modes interacting with qubits placed at positions $x^{(j)}$.
For the photons we assume a mode-independent group velocity $v$ and a linear dispersion relation, which is the relevant approximation as long as the bandwidth of the system is limited around the qubits' transition frequencies. We also separate the right- and left- moving photons, by introducing, for each of the $M$ modes indexed by  $m\in \{1,M\}$,  the corresponding field operators $\hat{\psi}_{m}^{(\rm r)}(x_{m})$, and $\hat{\psi}_{m}^{(\rm l)}(x_{m})$ ($x_m$ denotes the coordinate along the mode $m$).
The free-photon Hamiltonian then reads
\begin{equation}
H_{\rm ph} = \sum_{m=1}^{M}\int dx_{m}\left[\hat{\psi}_{m}^{(\rm r)+}(x_{m}) (-iv\partial_{x_{m}})\hat{\psi}_{m}^{(\rm r)}(x_{m}) + \hat{\psi}_{m}^{(\rm l)+}(x) (iv\partial_{x_{m}})\hat{\psi}_{m}^{(\rm l)}(x)\right] .
\end{equation}
For the qubit Hamiltonian we take
\begin{equation}
H_{\rm q}  = \sum_{j}\hbar(\omega_{10}^{(j)} - i \Gamma^{(j)} , /2)\hat{\sigma}^{(j)}_{11},
\end{equation}
where the two levels of the qubits are denoted by $|0\rangle ^{(j)}$ and $|1\rangle ^{(j)}$, $\omega_{10}^{(j)}$ is the interlevel frequency separation, $\hat{\sigma}_{st}=|s\rangle^{(j)(j)}\langle t|$, $s,t\in\{0,1\}$, and $\Gamma ^{(j)}$ describes the decay of the qubit $j$ due to other de-excitation channels than the ones provided  by the modes $m$ \cite{lukin}.
Finally, for the coupling between the qubits and the modes $m$ we take a rotating-wave approximation Hamiltonian
\begin{equation}
H_{\rm int} = \sum_{j} \int dx_{m} g^{(j)} \delta (x^{(j)})\left [\hat{\psi}_{m}^{(r)}(x)+\hat{\psi}_{m}^{(l)}(x)\right]\hat{\sigma}_{10}^{(j)} + h.c.
\end{equation}
The total Hamiltonian $H=H_{\rm q}+H_{\rm ph}+H_{\rm int}$ describes a quantum network which, as we show below, can be solved exactly. The general form of the solution of the Schr\"odinger equation can be written as
\begin{equation}
|\Psi (t)\rangle = \sum_{m=1}^{M} \int dx_{m}
\left[\phi_{m}^{\rm (r)}(x_{m},t)\hat{\psi}_{m}^{\rm (r)+}(x_{m}) +
\phi_{m}^{\rm (l)}(x_{m},t)\hat{\psi}_{m}^{\rm (l)+}(x_{m})\right]
|{\rm vac}\rangle + \sum_{j=1}^{N}q^{(j)}(t)\hat{\sigma}_{10}^{(j)}|{\rm vac}\rangle ,
\end{equation}
where $|{\rm vac}\rangle$ is the vacuum of all the qubits and modes, $q^{(j)}(t)$'s are the qubit's amplitudes, and $\phi_{m}^{\rm (r,l)}(x_{m},t)$ are the wavefunctions of the right-and left- propagating photons.

We then search for stationary solutions satisfying the time-independent Schr\"odinger equation
$H |\Psi (t)\rangle = \epsilon |\Psi (t)\rangle$ by using the parametrization $\phi_{m}^{\rm (r,l)} (x_{m},t) =
\exp (-i\epsilon t/\hbar ) \phi_{m}^{\rm (r,l)} (x_{m})$, $q^{(j)}(t) = \exp (-i \epsilon t/\hbar )q^{(j)}$, with
\begin{eqnarray}
\phi_{m}^{\rm (r)} (x_{m}) &=& e^{ikx_{m}}\sum_{j} \tilde{a}^{(j)}_{m}\Theta (x^{(j)}-x)\Theta (x - x^{(j-1)}) ,\\
\phi_{m}^{\rm (l)} (x_{m}) &=& e^{-ikx_{m}}\sum_{j} \tilde{b}^{(j)}_{m}\Theta (x^{(j)}-x)\Theta (x - x^{(j-1)}) .
\end{eqnarray}

With this parametrization, the Schr\"odinger equation results in $\epsilon = vk$ and
\begin{eqnarray}
g^{(j)}q^{(j)}&=& i v(\tilde{a}^{(j+1)}_{m} - \tilde{a}^{(j)}_{m})e^{ikx^{(j)}_{m}} , \nonumber\\
g^{(j)}q^{(j)}&=& - i v(\tilde{a}^{(j+1)}_{m} - \tilde{a}^{(j)}_{m})e^{-ikx^{(j)}_{m}} , \nonumber \\
\epsilon q^{(j)} &=& \hbar(\omega_{eg}^{(j)} - i\Gamma^{(j)} /2) q^{(j)} + \frac{g^{(j)}}{2}\sum_{m}\left[
\tilde{a}_{m}^{(j)}e^{ikx^{(j)}_{m}}+\tilde{a}_{m}^{(j+1)}e^{ikx^{(j)}_{m}}+
\tilde{b}_{m}^{(j)}e^{-ikx^{(j)}_{m}}+\tilde{b}_{m}^{(j+1)}e^{-ikx^{(j)}_{m}}\right] . \nonumber
\end{eqnarray}

A transparent way to obtain a solution for this system of equations is to use the method of  transfer matrices. We first define amplitudes containing phase factors $\exp (ikx_{m})$  in each of  the intervals between
consecutive nodes,
$a_{m}^{(j+1)}(x)= \tilde{a}_{m}^{(j+1)}\exp (ikx_{m})$, $b_{m}^{(j+1)}(x)= \tilde{b}_{m}^{(j+1)}\exp (-ikx_{m})$, where  $x_{m}\in [x_{m}^{(j)},x_{m}^{(j+1)}]$. Then we notice that between two consecutive nodes $j$ and $j+1$ the fields just pick up some phase factors: it is then natural to concentrate first on the scattering
process around one node $j$. Introducing a column vector $X^{(j)}$, $X^{(j)}_{1,2m-1}(x_{m}) = a_{m}^{(j)}(x_{m})$, and $X^{(j)}_{1,2m}(x_{m}) = b_{m}^{(j)}(x_{m})$ we obtain the result of scattering around the node $j$ in a form which generalizes previous theoretical results \cite{shen,solano}
\begin{equation}
X^{(j+1)}(x^{(j)}_m)=\left( 1 + \frac{A^{(j)}}{\gamma^{(j)}}\right)X^{(j)}(x^{(j)}_{m}) .
\end{equation}
Here $A$ is a $M \times M$ matrix with 1 on the odd rows and -1 on the even rows, $A_{l,s} = (-1)^{l+1}$, and a useful property is $A^2=0$. The quantity $\gamma^{(j)}$ is defined as
\begin{equation}
\gamma^{(j)} = \frac{\Gamma^{(j)}}{2}\frac{v}{g^{(j)2}} - i (E_{k}-\hbar\omega_{eg}^{(j)})\frac{v}{g^{(j)2}} .
\end{equation}
Between two consecutive nodes $j$ and $j+1$, the output amplitudes $a_{m}^{(j+1)}(x^{(j)}_{m})$, $b_{m}^{(j+1)}(x^{(j)}_{m})$ of the node $j$ are related to the input amplitudes $a_{m}^{(j+1)}(x^{(j+1)}_{m})$, $b_{m}^{(j+1)}(x^{(j+1)}_{m})$ of the node $j$ by
$a_{m}^{(j+1)}(x^{(j+1)}_{m})= a_{m}^{(j+1)}(x^{(j)}_{m})\exp [ik_{m}(x^{(j+1)}_{m}-x^{(j)}_{m})+ i\varphi^{(j,j+1)}_{m}]$, $b_{m}^{(j+1)}(x^{(j+1)}_{m})= b_{m}^{(j+1)}(x^{(j)}_{m})\exp [-ik_{m}(x^{(j+1)}_{m}-x^{(j)}_{m})- i\varphi_{m}^{(j,j+1)}]$. Here $\varphi_{m}^{(j,j+1)}$ represents an additional phase accumulated in mode $m$ due for example to the presence of a material which modifies the speed of light. One can introduce a diagonal "phase" matrix $\Phi^{(j,j+1)}$ which takes care of these effects: $\Phi^{(j+1,j)}_{l,s} = \exp [(-1)^{l+1}ik(x^{(j+1)}_{m}-x^{(j)}_{m})+ (-1)^{l+1}\varphi^{(j,j+1)}_{m}]\delta_{l,s}$. As a result, the amplitudes after the node $j+1$ can be obtained in the form of a product
\begin{equation}
X^{(j+2)}(x_{m}^{(j+1)}) = \left(1 - \frac{A}{\gamma^{(j+1)}}\right)\Phi^{(j+1,i)}\left(1 - \frac{A}{\gamma^{(j)}}\right)\Phi^{(j,i-1)} ... \left(1 - \frac{A}{\gamma^{(j-n)}}\right) X^{(j-n)}(x_{m}^{(j-n)}) .
\end{equation}

\begin{figure}[t]
\begin{center}
  \includegraphics[width=7cm]{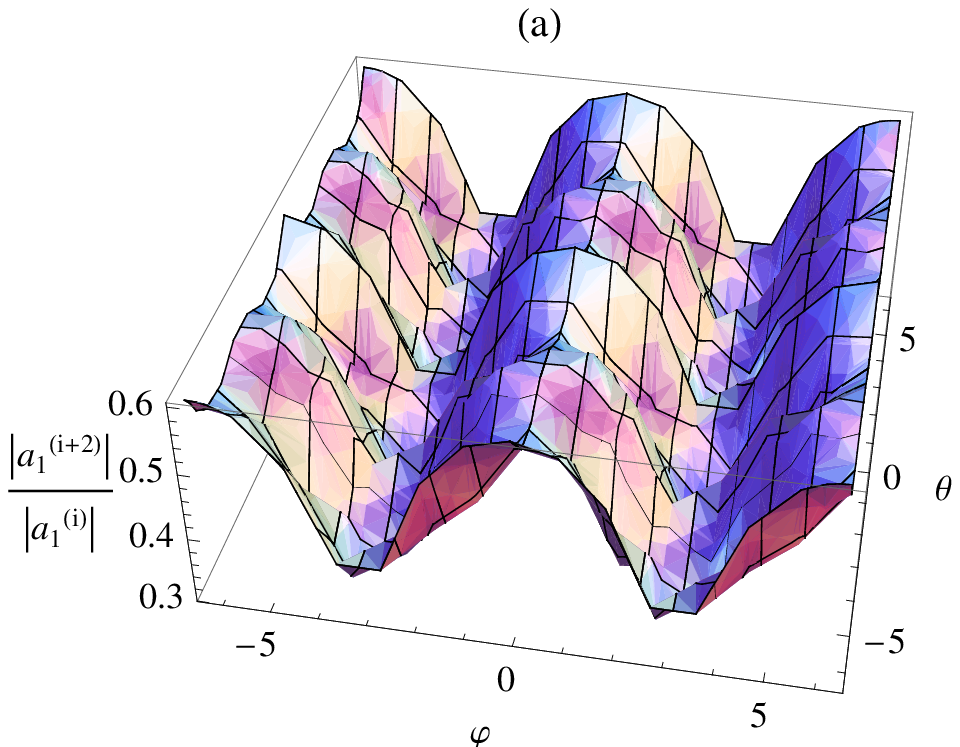}
  \includegraphics[width=7cm]{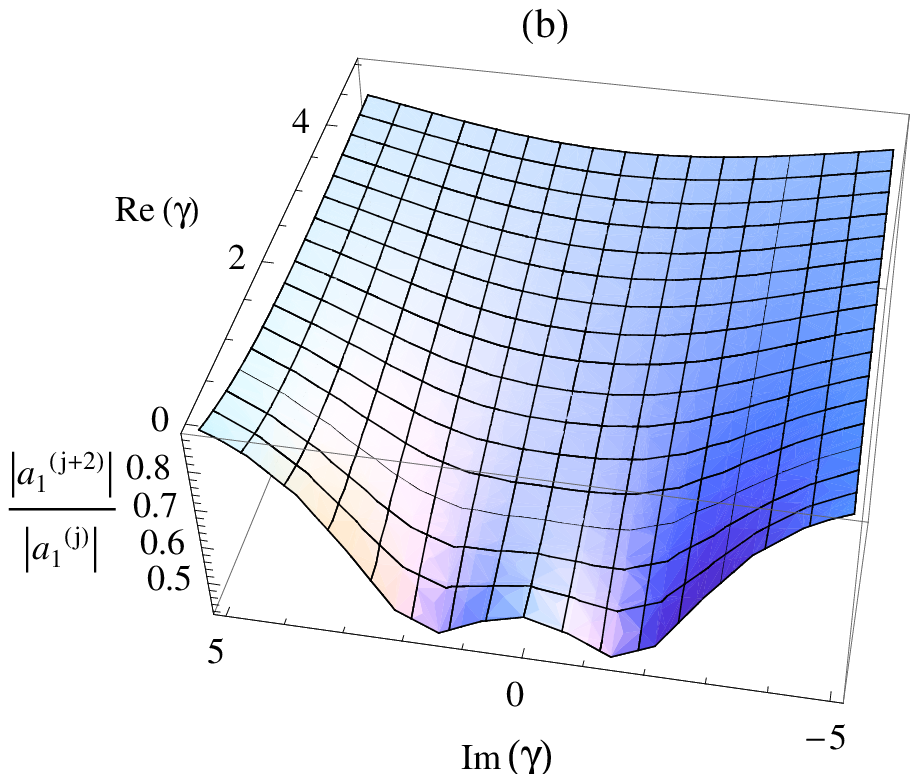}
\end{center}
\caption{We plot the absolute value of the transmission in the mode 1 as a function of (a) the phases $\varphi$, $\theta$ for $\gamma =1$, and (b) the real and imaginary parts of $\gamma$ (for $\theta = \pi/2$ and $\varphi = 0$). }
         \label{fig1}
\end{figure}

\section{Interferometer}

Now we employ this general formalism for the situation in which there are only two qubits, say $j$ and $j+1$. This situation is obtained by taking $x^{(n)}=-\infty$ for $n<j$ and $x^{(n)}=\infty$ for $n>j+1$. We also consider only two modes, $m=1,2$ and assume that the paths
between $j$ and $j+1$ of the two modes are identical, $\theta \stackrel{\rm not}{=}k(x^{(j+1)}_{m}-x^{(j)}_{m})$ and an additional phase difference $\varphi^{(j,j+1)}_{2}\stackrel{\rm not}{=}\varphi$ is produced in the second mode. We assume a single input field $a_{1}\neq 0$ at the input  of the resulting interferometer, creating four scattered components, $a_{1}^{(j+2)}(x_{1}^{(j+1)})$, $a_{2}^{(j+1)}(x_{2}^{(j+1)})$, $b_{1}^{(j)}(x_{1}^{(j)})$, and $b_{2}^{(j)}(x_{2}^{(j)})$. For simplicity the qubits are presumed identical, $\gamma^{(j)} = \gamma^{(j+1)} = \gamma$. Then the output amplitudes for this device   are obtained as
\begin{eqnarray}
a_{1}^{(j+2)}(x_{1}^{(j+1)})&=&e^{i\theta}\frac{-e^{i\varphi} + e^{2i(\varphi + \theta  )} + e^{i(\varphi + 2\theta )}-(1 + \gamma)^{2}}{4e^{i(\varphi + 2\theta )}\cos^{2}\varphi/2 - (2+\gamma )^{2}}a_{1}^{(j)}(x_{1}^{(j)}) ,\label{a1}\\
a_{2}^{(j+1)}(x_{2}^{(j+1)})&=&e^{i\theta}\frac{(1+e^{i\varphi})\left[-1-\gamma + e^{i(\varphi +2\theta )}\right]}{4e^{i(\varphi + 2\theta )}\cos^{2}\varphi/2 - (2+\gamma )^{2}}a_{1}^{(j)}(x_{2}^{(j)}) ,\\
b_{1}^{(j)}(x_{1}^{(j)})&=&\frac{2-2e^{i(\varphi + 2\theta )} + \gamma (1+e^{2i\theta} )}{4e^{i (\varphi + 2\theta )}\cos^{2}\varphi/2 - (2+\gamma )^{2}}a_{1}^{(j)}(x_{1}^{(j)}) ,\\
b_{2}^{(j)}(x_{2}^{(j)})&=&-\frac{-2 + e^{2i\theta } + e^{2i (\varphi + \theta )} - \gamma \left[1 + e^{i(\varphi + 2\theta)}\right]
}{4e^{i (\varphi + 2\theta )}\cos^{2}\varphi/2 - (2+\gamma )^{2}}a_{1}^{(j)}(x_{2}^{(j)}) .
\end{eqnarray}
The phase and amplitude information of the transmitted and reflected components $a_{1}^{(j+1)}$, $a_{2}^{(j+1)}$, respectively $b_{1}^{(j)}$, $b_{2}^{(j)}$ can be detected by vector network analyzer techniques. In Fig. \ref{fig1} we present the amplitude of the transmission in the mode 1, as obtained from Eq. (\ref{a1}). Fig. \ref{fig1} (a) demonstrates that this system behaves indeed as an interferometer, with the transmission displaying interference fringes ($2\pi$-periodicity as a function of $\varphi$ and $\pi$-periodicity as a function of $\theta$). For this figure we took $\gamma=1$, which corresponds to resonance and decay rate $\Gamma$ and coupling $g$ easily accessible experimentally \cite{solano}.
Also, from Fig. \ref{fig1} (b) we see that the transmission at large detunings approaches unity: this is expected, since in this case the qubits will absorb very little energy.

\section{Conclusion}

In conclusion, we propose a new concept for an interferometer based on the fluorescence of two qubits coupled by two quasi one-dimensional modes. 

\section{Acknowledgement}

This work was supported by the Academy of Finland (Acad. Res. Fellowship 00857, and projects 129896, 118122, and 135135).

\end{document}